# Comparative Analysis of Epileptic Seizure Prediction: Exploring Diverse Pre-Processing Techniques and Machine Learning Models

[1]Md. Simul Hasan Talukder; [2]Rejwan Bin Sulaiman

[1]Bangladesh Atomic Energy Regulatory Authority, Bangladesh; Email: simulhasantalukder@gmail.com

[2]Northumbria University, UK; Email: Rejwan.sulaiman@northumbria.ac.uk

**Abstract:** Epilepsy is a prevalent neurological disorder characterized by recurrent and unpredictable seizures, necessitating accurate prediction for effective management and patient care. Application of machine learning (ML) on electroencephalogram (EEG) recordings, along with its ability to provide valuable insights into brain activity during seizures, is able to make accurate and robust seizure prediction an indispensable component in relevant studies. In this research, we present a comprehensive comparative analysis of five machine learning models - Random Forest (RF), Decision Tree (DT), Extra Trees (ET), Logistic Regression (LR), and Gradient Boosting (GB) - for the prediction of epileptic seizures using EEG data. The dataset underwent meticulous preprocessing, including cleaning, normalization, outlier handling, and oversampling, ensuring data quality and facilitating accurate model training. These preprocessing techniques played a crucial role in enhancing the models' performance. The results of our analysis demonstrate the performance of each model in terms of accuracy. The LR classifier achieved an accuracy of 56.95%, while GB and DT both attained 97.17% accuracy. RT achieved a higher accuracy of 98.99%, while the ET model exhibited the best performance with an accuracy of 99.29%. Our findings reveal that the ET model outperformed not only the other models in the comparative analysis but also surpassed the state-of-the-art results from previous research. The superior performance of the ET model makes it a compelling choice for accurate and robust epileptic seizure prediction using EEG data.

## 1. Introduction

Epilepsy stands as a complex and chronic neurological disorder that affects millions of individuals worldwide, manifesting through recurrent and unpredictable seizures [1]. It occurs due to abnormal chemical changes in the brain, which, in turn, lead to altered electrical functioning within our neural processes [2]. It can be generalized, involving the entire brain, or focal (partial), affecting only a specific part of the brain [3-4]. According the report of WHO, Epilepsy is a significant global health concern, affecting around 50 million people worldwide. Each year, an estimated 5 million new epilepsy cases are diagnosed globally, with higher rates observed in low- and middle-income countries (up to 139 per 100,000) compared to high-income countries (49 per 100,000) [5]. During a seizure, a person may experience convulsions, loss of consciousness, repetitive movements, unusual sensations, or even behavioral changes, leading to potential accidents and injuries, especially if they occur during activities like driving, swimming, or operating heavy machinery. The severity and frequency of seizures can vary significantly from person to person. So, to optimize patient care and enhance their quality of life, early detection and accurate prediction of these seizures have become crucial. In this pursuit, Electroencephalogram (EEG) recordings have emerged as invaluable tools, offering insights into brain activity patterns during seizure events and establishing themselves as vital components in seizure prediction studies [6]. It measures the electrical activity generated by the brain using electrodes placed on the scalp. In the context of epilepsy, EEG is particularly useful because it can detect abnormal patterns of electrical activity that are characteristic of seizures. During an epileptic seizure, there are distinct changes in the brain's electrical activity that can be observed on the EEG recording. These changes are called epileptiform discharges, and they typically appear as sharp spikes or slow waves. The location, frequency, and pattern of these discharges can provide valuable information to help classify the type of epilepsy and determine the most appropriate treatment [7]. The problem is that due to burden of the patients, the doctor and radiologist become tiresome and sometime misdiagnosis appears. That's why driven by the urgent need to improve the management and treatment of neurological disorders, seizure prediction has become a focal point of intense research and exploration within the medical sciences. Among the various seizure types, epileptic seizures stand out due to their unpredictable and irregular patterns, presenting unique challenges for precise prediction. As our understanding of brain dynamics and computational methodologies continues to advance, it becomes increasingly apparent that sophisticated pre-processing techniques and machine learning models are essential to unlock the full potential of seizure prediction. These techniques play a pivotal role in extracting relevant information from raw EEG data and enhancing the quality of data available for predictive models.  By effectively combining advanced pre-processing techniques with powerful machine learning models, researchers and healthcare practitioners aim to develop accurate and reliable seizure prediction systems. The ultimate goal is to aid medical professionals in making well-informed decisions, providing timely interventions, and significantly improving the overall management and treatment of neurological disorders. Through this pursuit, the lives of epilepsy patients can be positively impacted, offering them better control over their condition and an improved quality of life.

In this research, we conduct a thorough comparative analysis of various machine learning models such as five traditional machine learning models, namely Random Forest (RF), Decision Tree (DT), Extra Trees (ET), Logistic Regression (LR), and Gradient Boosting (GB) to predict epileptic seizures using EEG data. The dataset consists of recordings from 500 individuals, each representing brain activity during different conditions, including eyes open, eyes closed, tumor presence, healthy brain regions, and instances of epileptic seizure events. Before model training, the dataset undergoes a meticulous preprocessing stage to ensure data cleanliness and robustness. The contributions of this research are as follow as-

- Unwanted data cleaning.
- Outlier detection of the dataset and managed properly.
- Introduction of random over sampling.
- Data normalization
- Comparative analysis with five ML models, namely RF, DT, ET, LR, and GB.
- Enhancing the performance of traditional ML models.
- Proposing highly accurate ET model for elliptical seizure prediction.

In this article, Section 2 represents literature review. The methodology, result analysis and discussion are described in the sections 3 and 4 respectively.  Conclusion and future work are illustrated in section 5.

## 2. Literature review

Different researches are currently underway to improve Epileptic Seizure detection from various perspectives. Some studies focus on dataset preprocessing, others on machine learning algorithms, and some on both aspects. Notably, the MIT-CHB and UCI epileptic seizure recognition datasets are publicly available for research purposes [8].

In a recent study [9], an Artificial Neural Network (ANN) approach with a specialized cost function was developed for EEG epileptic seizure detection. The authors achieved a high accuracy with an f1-score of 86%. This investigation also addressed the issue of imbalanced datasets using the CHB-MIT dataset. Our work has emphasized the UCI Epileptic Seizure dataset for early detection of epileptic seizures. AA Rahman et. al. [10] conducted a comparative analysis of Support Vector Machine (SVM), Random Forest (RF), and Multi-Layer Perceptron (MLP) models, both with and without hyperparameter tuning, achieving the best accuracy of 97.86% with SVM. In another study [11], Principal Component Analysis (PCA) with Genetic Algorithm-based Machine Learning (ML) approach was used for the binary classification of epileptic seizures from the UCI EEG dataset. The Extra Tree classifier outperformed other traditional ML models with hyperparameter tuning. Anwer Mustafa Hilal et al [12] proposed an intelligent deep canonical sparse autoencoder-based epileptic seizure detection and classification (DCSAE-ESDC) model using the UCI dataset. Their technique showed superior performance compared to existing techniques, achieving maximum accuracies of 98.67% and 98.73% under binary and multi-classification, respectively. K. Nanthini [13] proposed an LSTM model for detecting and predicting seizures on the UCI Database, achieving a remarkable 99% training and testing accuracy using Long Short-Term Memory networks. M Mahapatra et al [14] introduced an Autoencoder-Based Deep Neural Architecture for binary classification of epileptic seizures. The proposed architecture achieved a recognition accuracy of 92.47% on this task, with a comparative study against other standard neural models like deep neural networks (DNN) and CNN, as well as machine learning models such as logistic regression (LR), random forest (RF), and K-Nearest Neighbors (KNN). More recently, G. Xu [15] proposed a one-dimensional convolutional neural network-long short-term memory (1D CNN-LSTM) model for automatic recognition of epileptic seizures on the UCI dataset. The proposed method achieved high recognition accuracies of 99.39% and 82.00% on the binary and five-class epileptic seizure recognition tasks, respectively. Moreover, KM Alalayah et al [16] used the Epileptic Seizure Recognition UCI dataset for early detection of epileptic seizures. They applied Principal Component Analysis (PCA) and t-distributed stochastic neighbor embedding (t-SNE), along with K-means clustering + PCA and K-means clustering + t-SNE techniques, to reduce the dimensions of the features and identify important features for epilepsy. The authors employed various classifiers, including extreme gradient boosting, K-nearest neighbors (K-NN), decision tree (DT), random forest (RF), and multilayer perceptron (MLP), to classify the processed data. The MLP classifier with PCA + K-means achieved an accuracy of 98.98%, precision of 99.16%, recall of 95.69%, and F1 score of 97.4%.

From the literature review, it is evident that traditional machine learning models can achieve performance up to 98.98% on the UCI epileptic seizure recognition dataset, where the 1D CNN + LSTM models also perform well. While deep learning models can achieve high accuracy, they require substantial computational time and complexity. To enhance the classification performance of traditional machine learning models, this study embraced preprocessing techniques and employed the Extra Tree classifier, achieving promising results.

## 3. Materials and Methodology

The materials and methodology of this study are pointed out in this section step by step.

### 3.1. Dataset collection

We have collected dataset from UCI machine learning repository [17]. The dataset is a pre-processed and re-structured/reshaped version of a very commonly used dataset featuring epileptic seizure detection. It consists of EEG recordings collected from 500 individuals, each represented as a time-series with 4097 data points. To facilitate seizure prediction, the original time-series data was divided and shuffled into 23 chunks, each containing 178 data points representing 1 second of EEG recording. In total, the dataset contains 11500 samples, where each sample is a chunked time-series of 178 data points, and the last column represents the corresponding label y with values {1, 2, 3, 4, 5} indicating different states during recording depicted in Figure 1. The class distribution is shown in Table 1.

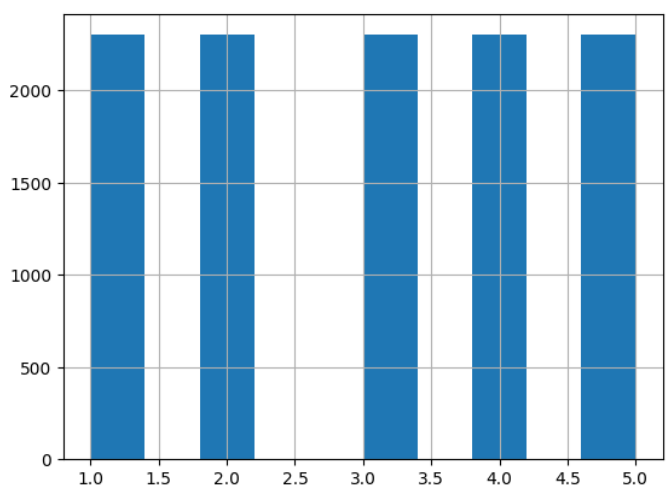

**Figure 1**. Data Distribution.

**Table 1**. Levels description of the classes.

| Label | Meanings |
|---|---|
| 5 | Eyes open |
| 4 | Eyes closed |
| 3 | Recording from the healthy brain area (EEG activity without tumor presence) |
| 2 | Recording from the area with tumor presence (EEG activity with tumor) |
| 1 | Recording of seizure activity (EEG activity during epileptic seizure) |

For simplification, binary classification is performed in our work, where subjects in class 1 (EEG during epileptic seizure) are differentiated from the rest (classes 2, 3, 4, and 5, representing non-seizure activities).

### 3.2. Dataset pre-processing

The quality of data preprocessing significantly influences model performance greatly [18]. During the preprocessing stage, several essential steps were meticulously executed to ensure the integrity and reliability of the EEG Seizure Prediction Dataset. Firstly, a redundant column labeled "Unnamed" that contained garbage values was identified and promptly removed to clean the dataset. Subsequently, to facilitate fair comparisons between different features, the dataset underwent normalization, which brought all variables to a common scale. This normalization process allowed each attribute to contribute equally to the analysis and model training, preventing any dominance of specific features. Next, outlier detection was conducted to identify extreme values that could potentially distort model performance. A total of 8269 outliers were detected across the various variables. To address their influence, a robust approach was employed, wherein the outliers were replaced with the median value of their respective columns. This replacement strategy ensured that the data distribution remained consistent and unaffected by extreme values. Finally, to gain insights into the interrelationships among the variables, a correlation analysis was performed. The resulting correlations were visualized in a Figure 2., revealing potential patterns and dependencies within the dataset. By completing these preprocessing steps, the EEG Seizure Prediction Dataset has been refined, normalized, and outlier-handled, paving the way for further analysis and the development of accurate predictive models for epileptic seizure detection.

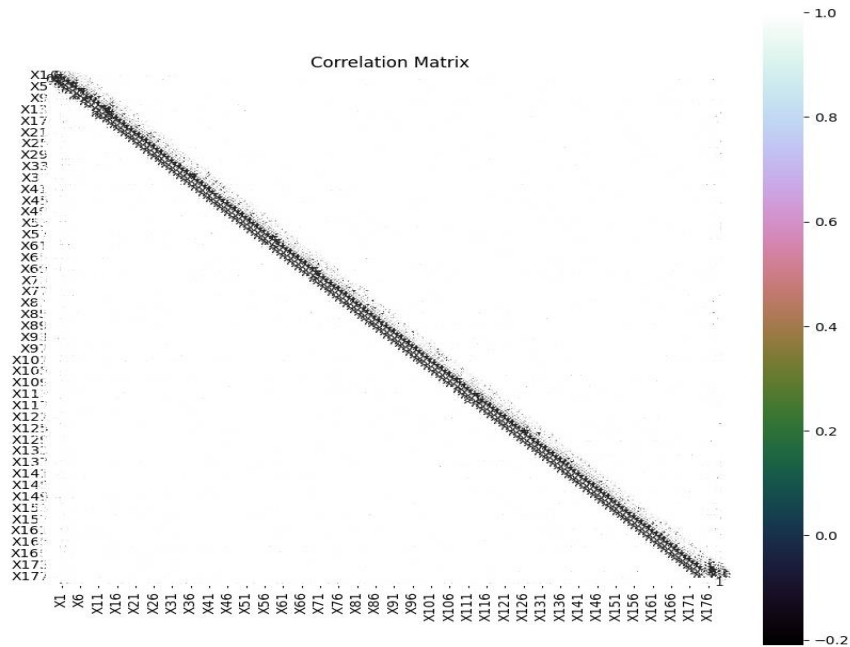

**Figure 2**. Data correlation matrix.

### 3.3.Proposed Model

In our study, the dataset underwent a series of essential preprocessing steps to ensure its suitability for accurate seizure prediction. Firstly, a thorough cleaning process was conducted, which involved removing the "Unnamed" column filled with garbage values. Next, to facilitate fair comparisons and eliminate scaling differences, the dataset was normalized, bringing all features to a standardized scale. To handle outliers effectively, we employed a robust approach, replacing the outlier values with the median of their respective columns. This step ensured that extreme values did not unduly influence the model's performance. In the context of seizure prediction, we recognized the importance of distinguishing between seizures and non-seizure activities. As such, a mapping process was implemented to designate non-seizure classes explicitly. To address class imbalance and enhance the model's ability to predict seizures accurately, we performed oversampling on the dataset, creating a balanced representation of the target classes. Before training and testing the models, we conducted Exploratory Data Analysis (EDA) to gain valuable insights into the dataset's characteristics and distributions. For model evaluation, we split the dataset into an 80:20 ratio for training and testing, respectively. We selected five traditional machine learning models, including RF, DT, ET, LR, and GB. Each model was trained and tested, and their performances were evaluated using various metrics, including confusion matrices, precision, recall, F1 score, accuracy, and the Receiver Operating Characteristic (ROC) curve. The whole procedures are presented in Figure 3.

After conducting the comparative analysis, we observed that the Extra Trees Classifier (ETC) consistently outperformed the other models across all evaluation parameters. Based on its superior performance, we have chosen the ETC as the proposed model for accurate prediction of elliptical seizures. By leveraging this model, we aim to provide an effective tool for early detection and intervention in epileptic patients, ultimately improving their quality of life and medical outcomes.

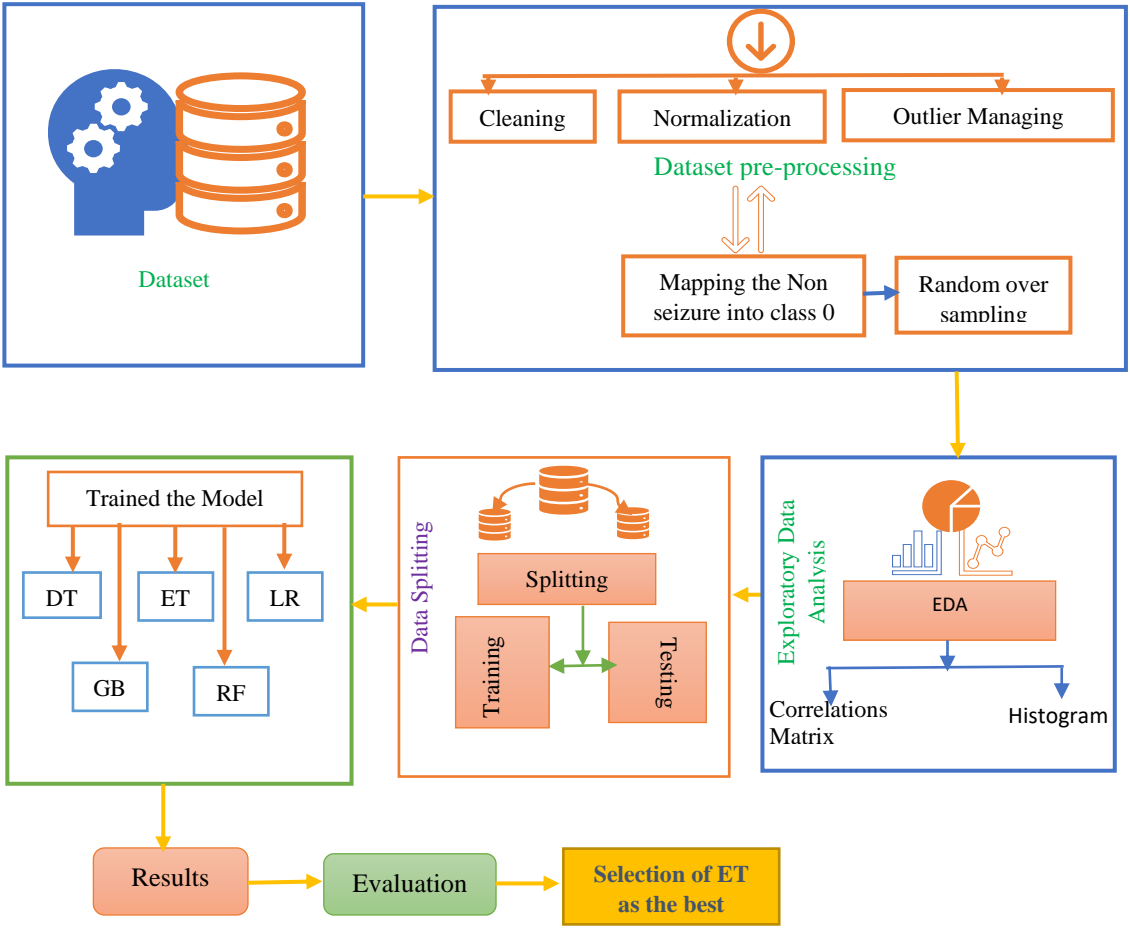

**Figure 3**. Proposed Methodology.

### 3.4. Model Evaluation

In our study, we conducted a comprehensive evaluation of the models using both quantitative metrics and visual analysis. For quantitative assessment, we utilized widely recognized performance metrics, including precision, recall, F1 score, and accuracy. These metrics provide valuable insights into the models' predictive capabilities and their ability to distinguish between different classes. The matrices are listed in Table 2.

**Table 2**. The matrices used in our study.

| Matrices name | Formulae |
|---|---|
| Precision [19] | $\frac{TP}{TP+FP}*100$ |
| Recall [19] | $\frac{TP}{TP+FN}*100$ |
| F1 Score [19] | $2 * \left(\frac{Precision * Recall}{Precision + Recall}\right) * 100$ |
| Accuracy [19] | $\frac{TP + TN}{TP + TN + FP + FN}$ |

Where TP denotes the true positive; TN is the true negative; FP denotes the false positive; FN denotes the false negative.

### 4. Result analysis and discussion

The performance metrics of several machine learning models were evaluated on the given dataset, showcasing diverse degrees of accuracy and effectiveness. The comprehensive results can be found in Table 3. Logistic Regression exhibited moderate performance, with an accuracy of 56.95% and all other metrics at 57%. This suggests that its linear decision boundary might not be well-suited for capturing complex patterns in the data. In contrast, Gradient Boosting and Decision Tree models achieved high accuracy (97.17%) and excellent precision, recall, and F1-scores (97%), indicating their ability to make accurate predictions. However, slight overfitting might be present. Random Forest and Extra Trees models outperformed others with impressive accuracy (98.99% and 99.29%, respectively) and near-perfect precision, recall, and F1-scores (99%). These ensemble methods demonstrated superior performance due to their ability to handle complex datasets effectively.

**Table 3.** Overall Performance matrices of the models.

| Models Name | Precision (%) | Recall (%) | F1-score (%) | Accuracy (%) |
|---|---|---|---|---|
| Logistic Regression | 57 | 57 | 57 | 56.95 |
| Gradient Boosting | 97 | 97 | 97 | 97.17 |
| Decision Tree | 97 | 97 | 97 | 97.17 |
| Random Forest | 99 | 99 | 99 | 98.99 |
| Extra Trees | 99 | 99 | 99 | 99.29 |

In the conducted study, confusion matrices were utilized to assess the performance of various machine learning models, particularly in scenarios involving class imbalances and varying degrees of misclassification impact. Figure 4 presents the confusion matrix, and based on a comparative analysis of the models, the following results were observed Among the evaluated models, the Extra Trees (ET) model demonstrated the highest performance, achieving the lowest number of misclassifications (26) out of 3679 instances. Random Forest (RF) and Decision Tree (DT) models also performed well, with relatively low misclassification counts (37 and 104, respectively). Gradient Boosting exhibited a moderate performance with 100 misclassifications. However, the Logistic Regression (LR) classifier showed the lowest effectiveness, with the highest misclassification count of 1583 instances.

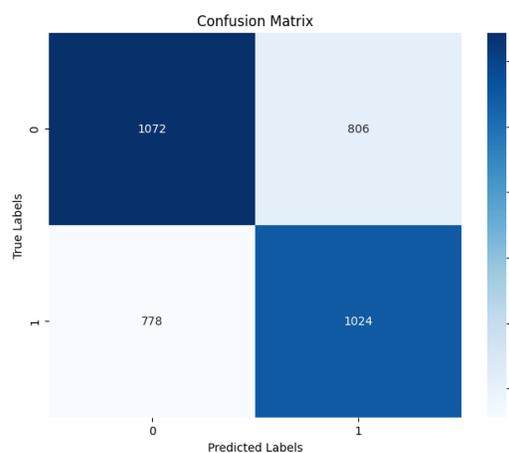

(a) Logistic Regression Classifier

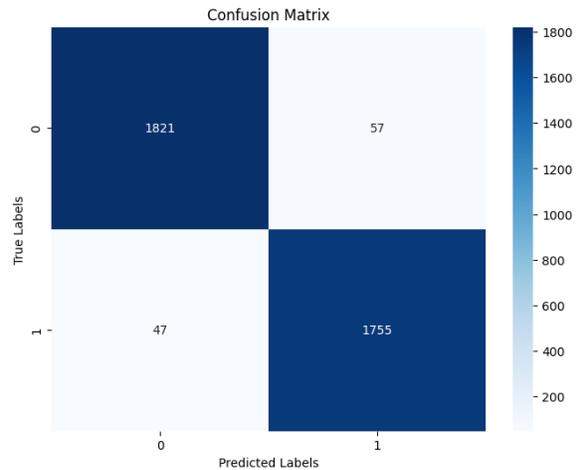

(b) Gradient Boosting classifier

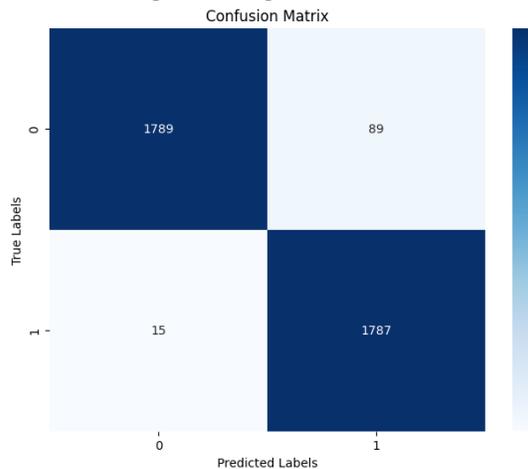

(c) Decision Tree classifier

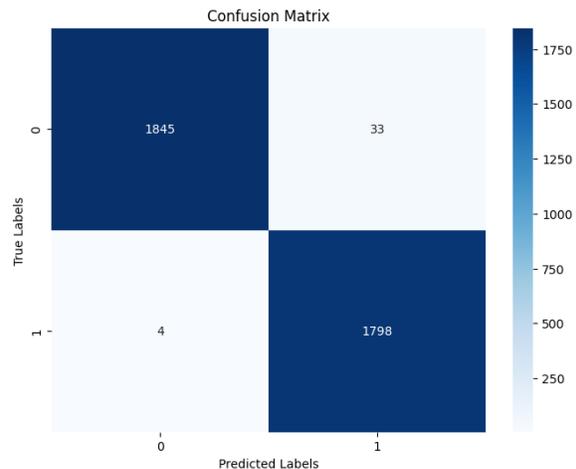

(d) Random Forest Classifier

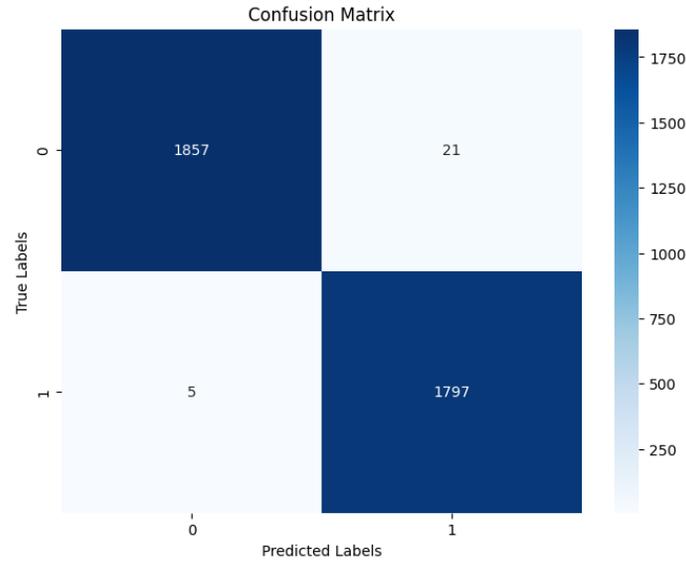

(e) Extra Tree Classifier

**Figure 4.** Confusion Matrix of the models.

In our evaluation, we used ROC curves, which graphically represent the ability of different machine learning models to distinguish between epileptic seizure and no seizure instances. Figure 5 illustrates the ROC curves for the following models: Logistic Regression, Gradient Boosting, Random Forest, Decision Tree, and Extra Tree classifier. The ROC-AUC values for these models were found to be 0.57, 0.97, 0.97, 0.99, and 0.99, respectively. The ROC curves demonstrate the trade-off between the true positive rate (TPR) and false positive rate (FPR) at various classification thresholds. Models with curves closer to the top-left corner indicate better performance, achieving higher TPRs with lower FPRs. Based on the ROC curves and AUC-ROC values, it is evident that the Extra Tree classifier achieved the highest AUC-ROC of 0.99, closely followed by Random Forest with an AUC-ROC of 0.99. Decision Tree and Gradient Boosting also performed well with AUC-ROC values of 0.97 each. However, the Logistic Regression model demonstrated significantly lower performance, indicated by its AUC-ROC of 0.57.

The ROC curve, confusion matrices, and various numerical analyses such as precision, recall, F1 score, and accuracy consistently indicate that the Extra Tree classifier is the most promising model for Epileptic seizure detection. This model achieves the highest AUC-ROC values, precision, recall, and accuracy, showcasing its effectiveness in recognizing epileptic seizures. The good balance between the true positive rate and false positive rate further supports its superior performance.

The success of the Extra Tree classifier can be attributed, in part, to the preprocessing steps undertaken in the study. Cleaning redundant and unnamed columns, normalizing the dataset, random oversampling, and replacing outliers with the median of the column played a crucial role in improving the model's performance. This combination of data preprocessing techniques helped create a more robust and accurate model.

Comparing the proposed model with previous studies, it is evident that some existing approaches did not utilize ensemble methods like the Extra Tree classifier. Additionally, they may have overlooked specific preprocessing steps that contributed to the overall performance improvement observed in this study. For instance, the models presented in [10] and [16] did not leverage the Extra Tree classifier, potentially missing out on the benefits of ensemble learning. Furthermore, while [11] did employ the Extra Tree classifier, it did not place emphasis on crucial data preprocessing steps such as outlier replacement with the median and random oversampling. These preprocessing techniques, as demonstrated in our study, play a pivotal role in enhancing model robustness and accuracy.

While the 1D CNN-LSTM model [15] achieved a higher accuracy of 99.39%, it may not provide a balanced precision, recall, and accuracy like the traditional models [12-16]. In contrast, the proposed model in this study achieved balanced values of precision, recall, accuracy, and F1 score, all around 99%, outperforming the other existing traditional machine learning models.

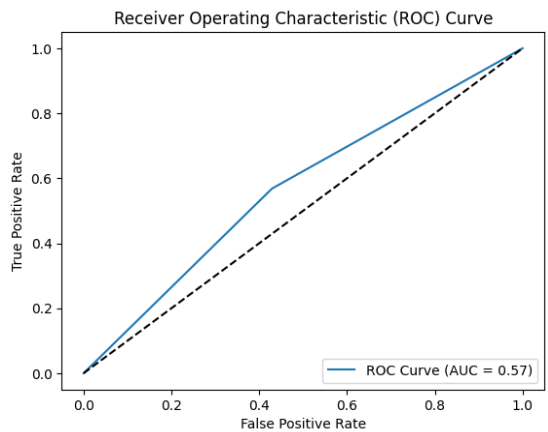

(a) Logistic Regression Classifier

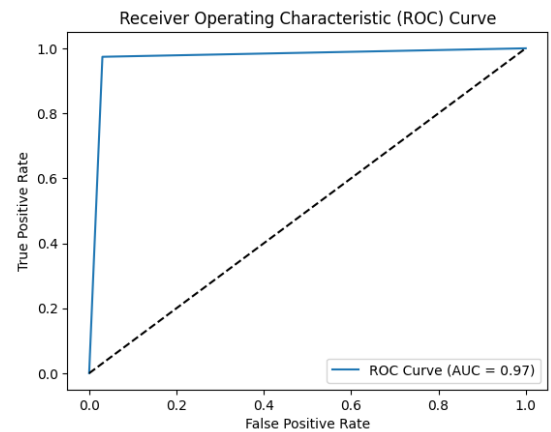

(b) Gradient Boosting classifier

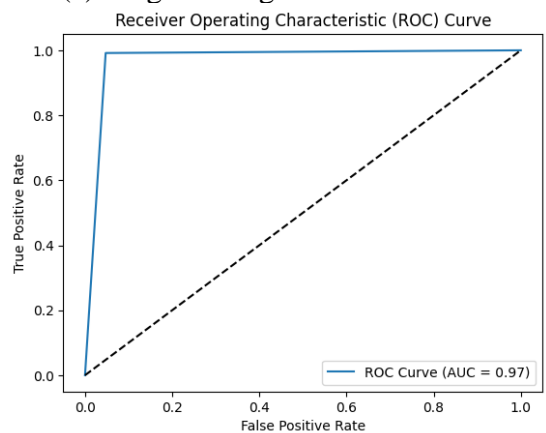

(c) Decision Tree classifier

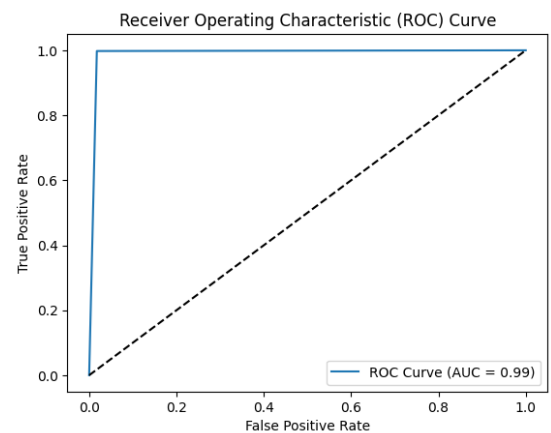

(d) Random Forest Classifier

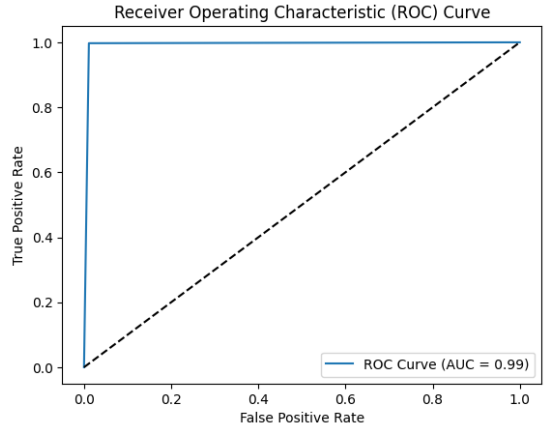

(e) Extra Tree Classifier

**Figure 5**. ROC curves of the models.

**Table 4.** Comparative analysis with the previous work.

| Models | | Accuracy (%) | Precision (%) | Recall (%) | F1 score (%) |
|---|---|---|---|---|---|
| DCSAE-ESDC [12] | | 98.67 | - | - | - |
| 1D CNN-LSTM [15] | | **99.39** | - | - | - |
| [16] * **Very recent work** | DWT and PCA using Random Forest | 97.96 | 99.10 | 94.41 | 97.41 |
| | DWT and t-SNE using Random Forest | 98.09 | 99.10 | 93.9 | 96.21 |
| | K-Means with PCA using MLP | 98.98 | 99.16 | 95.69 | 97.4 |
| Our proposed work | RF | 98.99 | 99 | 99 | 99 |
| | **Extra tree** | **99.29** | **99** | **99** | **99** |

## 5. Conclusion and Future Work

The comparative analysis of epileptic seizure prediction, exploring diverse pre-processing techniques, and machine learning models has provided valuable insights into the effectiveness of different approaches for this critical medical task. The study evaluated multiple pre-processing techniques, such as cleaning, outlier managing, normalization and random over sampling to prepare the epileptic seizure data for modeling. It was observed that the choice of pre-processing techniques significantly impacted the performance of the machine learning models. Various machine learning algorithms, including Logistic Regression, Gradient Boasting Decision Trees, Random Forest, and Extra tree models, were tested on the pre-processed data. Each model exhibited different strengths and weaknesses in predicting epileptic seizures. Among the machine learning models tested, extra tree model showed the most promising results with accuracy of 99.29%. It demonstrated robustness and high accuracy in seizure prediction due to its ensemble nature and ability to handle non-linear relationships in the data. It is the highest for traditional machine learning models as my knowledge.

However, it is essential to note that no single traditional ML model or pre-processing technique is universally superior for all scenarios. The choice of the most appropriate approach depends on factors such as the dataset size, quality, and the specific characteristics of the epileptic seizure data. In practical applications, considering the trade-offs between model complexity, interpretability, and computational resources is crucial. While Deep Learning models in [15] may achieve state-of-the-art performance, they often require more extensive datasets and computational power, which might not be feasible in all clinical settings. Therefore, Extra tree, with its balanced performance and relatively lower computational demands, could be a pragmatic choice for certain scenarios.

The comparative analysis provides a roadmap for future research in epileptic seizure prediction, highlighting the potential for further improvements in pre-processing techniques and machine learning algorithms. By refining the approaches used in this study and integrating domain-specific knowledge, we can develop more accurate and reliable seizure prediction models, ultimately contributing to better patient care and improved seizure management.

**Author Contributions: Md. Simul Hasan Talukder:** Implementation and manuscript drafting; **Rejwan Bin Sulaiman:** Review, correction and guidance.

**Funding:** There has no external funding.

**Data Availability Statement:** https://archive.ics.uci.edu/dataset/388/epileptic+seizure+recognition